%% file: preprint.tex
\documentclass[a4paper]{my}

\usepackage[latin1]{inputenc}
\usepackage{hyperref}
\usepackage{amsfonts,amsmath,cite}
\usepackage{pstricks,pst-node,multido}

\hypersetup{
    colorlinks=false,
    pdfborder={0 0 0},
    pdftitle={An optimal local approximation algorithm for max-min linear programs},
    pdfauthor={Patrik Floréen, Joel Kaasinen, Petteri Kaski, Jukka Suomela},
}

\suPerfalse

\input{fig-settings.tex}

\newcommand{\mysize}[1]{{\lvert #1 \rvert}}
\newcommand{\myv}[1]{\mathbf{#1}}
\newcommand{\myA}{\mathcal{A}}
\newcommand{\myG}{\mathcal{G}}
\newcommand{\Dk}{\Delta_K}
\newcommand{\Di}{\Delta_I}
\newcommand{\myaspacing}{a^{\vphantom{-}}}
\newcommand{\myotherVK}[1]{N(#1)}
\newcommand{\myotherVI}[2]{n(#1,#2)}

\theoremstyle{definition}

\newtheorem*{uremarks}{Remarks}

\begin{document}

\title[]{An optimal local approximation algorithm \\ for max-min linear programs}

\author{P. Floréen}{Patrik Floréen}
\author{J. Kaasinen}{Joel Kaasinen}
\author{P. Kaski}{Petteri Kaski}
\author{J. Suomela}{Jukka Suomela}
\address{%
    Helsinki Institute for Information Technology HIIT \newline
    Helsinki University of Technology and University of Helsinki \newline
    P.O. Box 68, FI-00014 University of Helsinki, Finland \newline
    \emph{E-mail addresses: }{\tt patrik.floreen@cs.helsinki.fi}, {\tt joel.kaasinen@cs.helsinki.fi}, \newline
    \phantom{\emph{E-mail addresses: }}{\tt petteri.kaski@cs.helsinki.fi}, {\tt jukka.suomela@cs.helsinki.fi}}

\thanks{This research was supported in part by the Academy of Finland, Grants 116547 and 117499, and by Helsinki Graduate School in Computer Science and Engineering (Hecse).}


\begin{abstract}
    \noindent We present a local algorithm (constant-time distributed algorithm) for approximating max-min LPs. The objective is to maximise $\omega$ subject to $A \myv{x} \le \myv{1}$, $C \myv{x} \ge \omega \myv{1}$, and $\myv{x} \ge \myv{0}$ for nonnegative matrices $A$ and $C$. The approximation ratio of our algorithm is the best possible for any local algorithm; there is a matching unconditional lower bound.
\end{abstract}

\maketitle

\section{Introduction}

In a \emph{max-min linear program} (max-min LP), the objective is to
\begin{equation}\label{eq:max-min-intro}
\begin{alignedat}{2}
    &\text{maximise } \ & {\textstyle\min_{k \in K} \myv{c}_k \myv{x}} & \\
    &\text{subject to } \ & A \myv{x} \,&\le\, \myv{1}, \\
    && \myv{x} \,&\ge\, \myv{0}
\end{alignedat}
\end{equation}
or, equivalently, to maximise $\omega$ subject to $A \myv{x} \le \myv{1}$, $C \myv{x} \ge \omega \myv{1}$, and $\myv{x} \ge \myv{0}$. The matrices $A$ and $C$ are nonnegative and sparse: each row $\myv{a}_i$ of $A$ has at most $\Di$ positive elements, and each row $\myv{c}_k$ of $C$ has at most $\Dk$ positive elements; here $\Di$ and $\Dk$ are constants.

Our work studies \emph{local algorithms} \cite{linial92locality,naor95what}, that is, distributed algorithms that complete in \emph{constant time} (constant number of synchronous communication rounds), independent of the size of the network. We assume that there is a network with one node $v \in V$ for each variable $x_v$, one node $i \in I$ for each constraint $\myv{a}_i \myv{x} \le 1$, and one node $k \in K$ for each objective $\myv{c}_k \myv{x}$. Nodes $v \in V$ and $i \in I$ are adjacent if $a_{iv}$ is positive, and nodes $v \in V$ and $k \in K$ are adjacent if $c_{kv}$ is positive. In a local algorithm for max-min LPs, the node $v \in V$ must choose the value $x_v$ based on the information that is available within its constant-radius neighbourhood in the network.

Max-min LPs are a generalisation of packing LPs. Direct applications of max-min LPs include various tasks of fair resource allocation, such as fair bandwidth allocation in a communication network and balanced data gathering in a wireless sensor network. An algorithm for approximating max-min LPs allows one to solve approximate mixed packing and covering problems \cite{young01sequential} as well. Special cases of mixed packing and covering LPs include finding an (approximate) solution to a nonnegative system of linear equations.

\subsection{Prior work}

For any $\epsilon>0$, there exists a local $(1+\epsilon)$-approximation algorithm for packing and covering LPs, assuming a bounded-degree graph and bounded coefficients \cite{kuhn05price,kuhn06price}. However, this is not the case with max-min LPs:
\begin{theorem}[Floréen et al.~\cite{floreen08tight}]\label{thm:inapprox}
    For any $\Di \ge 2$ and $\Dk \ge 2$, there exists no local approximation algorithm for the max-min LP problem with the approximation ratio ${\Di (1 - 1/\Dk)}$.
\end{theorem}

This theorem holds even in the following special cases: (i)~\emph{bipartite} max-min LPs, where each column of $A$ and each column of $C$ contains only one nonzero element, and (ii)~$0/1$ max-min LPs, where each element of $A$ and $C$ is either $0$ or $1$.

The approximation ratio ${\Di (1 - 1/\Dk)} + \epsilon$ for any positive $\epsilon$ can be achieved by a local algorithm for bipartite problems \cite{floreen08tight} and for a class of $0/1$ problems \cite{floreen08local}. However, for general max-min LPs, the best known local algorithm has been the \emph{safe algorithm} that achieves the factor $\Di$ approximation \cite{floreen08approximating,papadimitriou93linear}.

\subsection{Contribution}

The following theorem summarises the main contribution of this paper.
\begin{theorem}\label{thm:approx}
    For any $\Di \ge 2$, $\Dk \ge 2$, and $\epsilon > 0$, there exists a local approximation algorithm for the max-min LP problem with the approximation ratio ${\Di (1 - 1/\Dk) + \epsilon}$.
\end{theorem}

By Theorem~\ref{thm:inapprox}, the algorithm in Theorem~\ref{thm:approx} is optimal in the sense that no local algorithm can achieve a better approximation ratio. For the case $\Di = \Dk = 2$, the approximation ratio is $1 + \epsilon$, that is, there is a local approximation scheme.

\subsection{Definitions}

We now give a detailed definition of a max-min LP in a distributed setting. Let $\myG = (V \cup I \cup K, E)$ be a bipartite, undirected communication graph. The nodes $v \in V$ are called \emph{agents}, the nodes $i \in I $ are called \emph{constraints}, and the nodes $k \in K$ are called \emph{objectives}; the sets $V$, $I$, and $K$ are disjoint. Edges $e \in E$ are of the form $e = \{v, i\}$ or $e = \{v, k\}$ where $v \in V$, $i \in I$, and $k \in K$.

Let
$V_i = {\{ v \in V : \{v,i\} \in E \}}$, 
$V_k = {\{ v \in V : \{v,k\} \in E \}}$, 
$I_v = {\{ i \in I : \{v,i\} \in E\}}$, and 
$K_v = {\{ k \in K : \{v,k\} \in E\}}$ 
for all $i \in I$, $k \in K$, and $v \in V$.
We assume that $\mysize{V_i} \le \Di$ and $\mysize{V_k} \le \Dk$ for all $i \in I$ and $k \in K$ for some constants $\Di$ and $\Dk$.

A \emph{max-min linear program} associated with $\myG$ is defined as follows. Associate a variable $x_v$ with each agent $v \in V$, associate a coefficient $a_{iv} > 0$ with each edge $\{i,v\} \in E$, $i \in I$, $v \in V$, and associate a coefficient $c_{kv} > 0$ with each edge $\{k,v\} \in E$, $k \in K$, $v \in V$. The task is to
\begin{equation}
    \begin{aligned}
        &\text{maximise } & \omega(\myv{x}) \,=\, 
         \textstyle\min_{k \in K} \textstyle\sum_{v \in V_k} c_{kv} x_v & \\
        &\text{subject to } & 
 \textstyle\sum_{v \in V_i} a_{iv} x_v \,&\le\, 1, & \quad\forall\, &i \in I, \\
   && x_v \,&\ge\, 0, & \forall\, &v \in V .
    \end{aligned}\label{eq:max-min}
\end{equation}

The \emph{local input} of an agent $v \in V$ consists of the sets $I_v$ and $K_v$ and the coefficients $a_{iv}, c_{kv}$ for all $i \in I_v, k \in K_v$. The local input of a constraint $i \in I$ consists of $V_i$, and the local input of an objective $k \in K$ consists of $V_k$. In a local algorithm, there is a constant $D$ such that each agent $v \in V$ chooses the output $x_v$ based on the local inputs of the nodes within distance $D$ (in number of edges) from $v$ in the communication graph~$\myG$; the constant $D$ is the \emph{local horizon} of the algorithm.

Theorem~\ref{thm:inapprox} holds even if we assume that each node of the graph $\myG$ has a unique identifiers. We show that Theorem~\ref{thm:approx} holds even if we do not have unique identifiers. We merely assume \emph{port numbering} \cite{angluin80local}: each edge $\{s,t\}$ in $\myG$ has two natural numbers associated with it, the port number in $s$ and the port number in $t$.

\subsection{Overview of the algorithm}

We begin in \S\ref{sec:unfolding} by reminding that, in the port numbering model, we can without loss of generality focus on the case where the communication graph $\myG$ is a (countably infinite but locally finite) tree \cite{angluin80local,floreen08tight}.

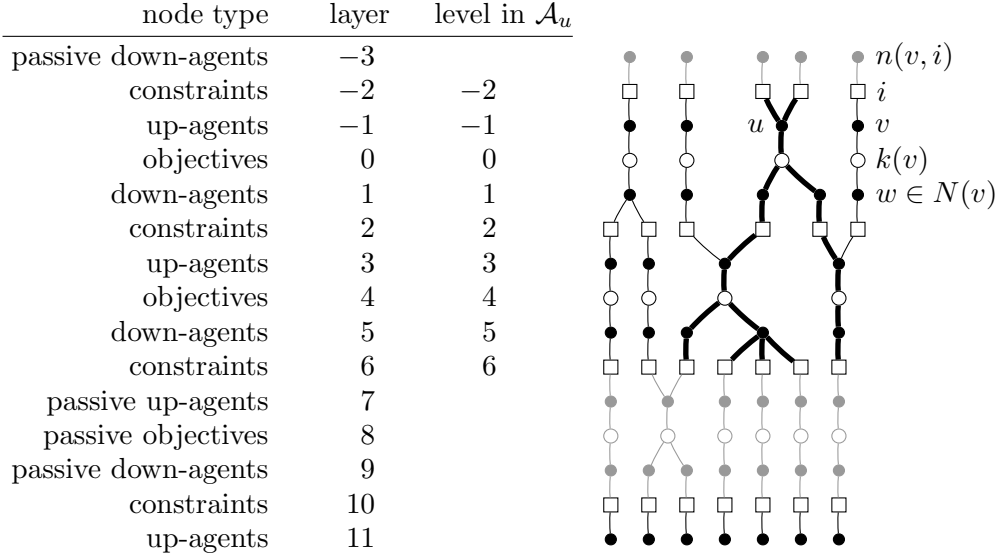
\begin{figure}
    \centering
    \input{fig-layers.tex}
    \caption{The graph $\myG$ (after the local transformations in \S\ref{sec:local-trans}) and the layers (see \S\ref{sec:analysis}).  We have chosen $R=3$ and hence $r=1$. Thick lines highlight the tree $\myA_u$ (see \S\ref{ssec:A_u}). There are several ways to choose the layers; nevertheless, if $u$ is an up-agent on layer $-1$ then the levels in $\myA_u$ necessarily coincide with the layers.}\label{fig:layers}
\end{figure}

In \S\ref{sec:local-trans}, we present a series of local transformations that simplify the structure of the problem. We show that with the objective of establishing Theorem~\ref{thm:approx}, it is sufficient to focus on the special case where $\mysize{V_i} = 2$, $\mysize{V_k} \ge 2$, $\mysize{K_v} = 1$, $\mysize{I_v} \ge 1$, and $c_{ku} = 1$ for all $i \in I$, $k \in K$, $v \in V$, $u \in V_k$. Figure~\ref{fig:layers} shows an example of a communication graph $\myG$ after the local transformations.

In \S\ref{sec:approx}, we present a local algorithm for this special case. In \S\ref{sec:analysis}, we prove that the output of the algorithm is a factor ${2 (1 - 1/\Dk)} + \epsilon'$ approximation. Put together, we have a local, factor ${\Di (1 - 1/\Dk)} + \epsilon$ approximation for general max-min LPs, for any $\epsilon > 0$.

The intuition behind the algorithm in \S\ref{sec:approx} is best understood if we study its analysis in \S\ref{sec:analysis}. In the analysis, it is convenient to assume that we have assigned a one-dimensional coordinate, \emph{layer}, to each node of the tree $\myG$; see Figure~\ref{fig:layers} for an example. When we assign the layers, we also partition the agents into \emph{up-agents} and \emph{down-agents}. We have alternatingly layers of up-agents, constraints, down-agents, and objectives. Each objective has exactly one adjacent up-agent ``above'' it, and at least one adjacent down-agent ``below'' it; hence the names ``up'' and ``down''.

We now assign every $R^\text{th}$ layer of objectives to be \emph{passive}, including the adjacent agents that set $x_v = 0$. Each agent $v \in V$ computes an upper bound $t_v$ of the optimum; see \S\ref{ssec:t_u}. Then we construct a solution for the active layers in a greedy manner, starting with a layer of passive up-agents and propagating information upwards until we reach the next layer of passive down-agents; see \S\ref{ssec:g}.

This is not yet an approximation for the original problem: while most objectives perform at least as well as in the global optimum, the passive objectives have utility $0$. By applying ideas from the shifting strategy \cite{baker94approximation,hochbaum85approximation}, we could consider $R$ possible choices for the locations of the passive layers. Then we could take averages over these to obtain a solution $\myv{y}$. In \S\ref{ssec:averaging} we show that $\myv{y}$ would indeed be a factor $R/(R-1)$ approximation.

There is one difficulty, however. We cannot assign the layers by a local algorithm in a globally consistent manner; in particular, we do not know whether a given agent is a down-agent or an up-agent. To overcome this, we consider both possible roles for each agent, up and down. For both roles, we compute a candidate solution by applying the shifting strategy. Finally we take the average of both candidate solutions. \mbox{This is the essence of \eqref{eq:x-def}.}

In \S\ref{sec:analysis}, we prove that this local approach yields a globally feasible solution, and the solution is within factor ${2 (1 - 1/\Dk)} + \epsilon'$ of the optimum. The constant $\epsilon' > 0$ can be made arbitrarily small by choosing a sufficiently large $R$.

Yet another basic hurdle implicitly overcome in the proof of Theorem~\ref{thm:approx} stems from \emph{underconstrained instances}\/: if there are several equally good solutions, one needs to choose between them in a globally consistent manner. Our definition of the values $g^+$ and $g^-$ in \mbox{\eqref{eq:g-recursion1}--\eqref{eq:g-recursion3}} addresses this by focusing on a particular extreme point. Each layer of down-agents chooses as large values $g^+$ as possible, without violating the constraints ``below'' them. Each layer of up-agents chooses as small values $g^-$ as possible, as long as the objectives ``below'' them meet the smoothed upper bounds $s_v$.

\section{Port numbering and unfolding}\label{sec:unfolding}

Our algorithm does not need to use any node identifiers; port numbering is sufficient. In the port numbering model, a local algorithm cannot distinguish between a short cycle and an infinitely long path. We can exploit this limitation to simplify the description of our local algorithm: we can assume that we have \emph{unfolded} all cycles of the graph $\myG$ \cite{angluin80local,floreen08tight}.

The definition of unfolding requires some preliminaries. A \emph{walk} of \emph{length} $\ell$ in a graph $\myG$ is a nonempty tuple $(u_0,e_1,u_1,e_2,u_2,\ldots,e_\ell,u_\ell)$ of alternating nodes and edges in $\myG$ such that, for all $j=1,2,\ldots,\ell$, the edge $e_j$ joins the nodes $u_{j-1}$ and $u_j$. The walk is said to \emph{start} at $u_0$ and \emph{end} at $u_\ell$. A walk is \emph{non-backtracking} if $e_{j-1}\neq e_j$ holds for all $j=1,2,\ldots,\ell$. A \emph{path} is a walk with no repeated nodes.

Let $\myG$ be a finite connected graph and let $r$ be a node of $\myG$. The \emph{unfolding} of $\myG$ \emph{rooted at} $r$ is the undirected simple graph $\myG'$ obtained as follows. The node set of $\myG'$ is the set of all non-backtracking walks in $\myG$ that start at $r$. Two nodes of $\myG'$ are joined by an edge iff one can be obtained from the other by appending exactly one edge and one node of $\myG$.

\pagebreak

We associate with each node of $\myG'$ a \emph{parent} node of $\myG$, namely the end-node of the walk. We also associate with each edge of $\myG'$ a \emph{parent} edge of $\myG$, namely the appended edge.

\begin{uremarks}\hfill
\begin{enumerate}
    \item The unfolding $\myG'$ is a tree.
    \item The unfolding $\myG'$ is finite iff $\myG$ is a tree; otherwise $\myG'$ is countably infinite.
    \item Any two unfoldings of $\myG$ rooted at different nodes are isomorphic. In what follows we refer to ``the'' unfolding of $\myG$ without specifying a particular root node.
    \item Assuming that the graph $\myG$ has port numbers associated with the ends of its edges, the unfolding $\myG'$ inherits the port numbering from the parent edges.
    \item Assuming that the graph $\myG$ has a max-min LP associated with it, the max-min LP associated with the unfolding $\myG'$ is defined by inheritance from the parent nodes and edges. In particular, the type of each node (agent, constraint, objective) is the type of the parent node, and the coefficients associated with the edges ($a_{iv}$,~$c_{kv}$) are inherited from the parent edges.
    \item Any two nodes of $\myG'$ with the same parent are related by an automorphism of $\myG'$. In particular, any deterministic local algorithm in the port numbering model must give the same output on any two nodes with the same parent. Any locally computable feasible solution of the max-min LP associated with $\myG'$ defines a feasible solution of the max-min LP associated with $\myG$, with the same utility.
    \item Any feasible solution of the max-min LP associated with $\myG$ defines, by inheritance, a feasible solution of the max-min LP associated with $\myG'$, with the same utility.
    \item Any locally computed feasible solution of $\myG'$ with utility at least $1/\alpha$ times the utility of any feasible solution of $\myG'$ yields an $\alpha$-approximation of the optimum of $\myG$.
\end{enumerate}
\end{uremarks}

\section{Local transformations}\label{sec:local-trans}

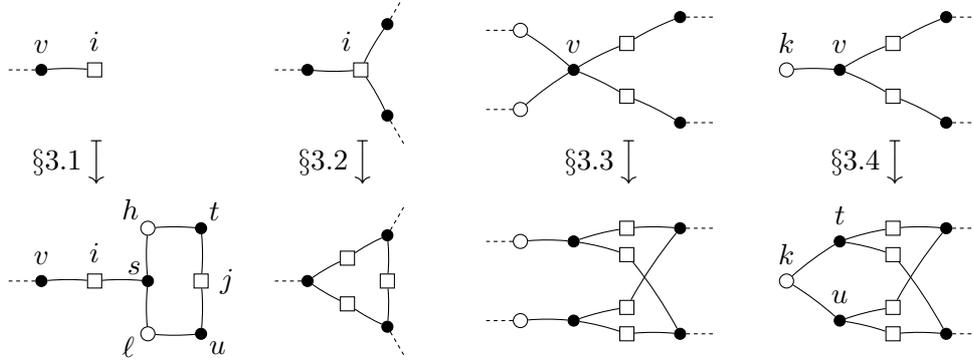
\begin{figure}
    \centering
    \input{fig-trans.tex}
    \caption{Local transformations in \S\ref{ssec:trans-vi-ge-2}--\S\ref{ssec:trans-vk-ge-2}.}\label{fig:trans}
\end{figure}

Consider an arbitrary max-min LP associated with the graph $\myG$. In this section we carry out a sequence of locally computable transformations, with the goal of arriving at a more structured max-min LP. The transformations are applied in the order of presentation, from \S\ref{ssec:trans-vi-ge-2} to \S\ref{ssec:trans-c01}. We describe each individual transformation in three parts:
\begin{enumerate}
    \item A description of the transformation.
    \item Mapping a solution of the transformed instance back to the original instance.
    \item Implications to approximability. We write $\omega(\cdot)$ for the utility of the original instance and $\omega'(\cdot)$ for the utility of the transformed instance.
\end{enumerate}
Appendix~\ref{app:trans-comp} presents the implementation of these transformations in the port numbering model. Figure~\ref{fig:trans} illustrates the transformations that modify the communication graph~$\myG$.

To avoid degenerate cases, we assume that each constraint and objective is adjacent to at least one agent, and every agent is adjacent to at least one constraint and at least one objective, that is, $\mysize{V_i} \ge 1$, $\mysize{V_k} \ge 1$, $\mysize{K_v} \ge 1$, and $\mysize{I_v} \ge 1$ for all $i \in I$, $k \in K$, $v \in V$. Indeed, isolated constraints can be deleted, isolated objectives force the optimum of \eqref{eq:max-min} to zero, non-contributing agents can be set to zero, and unconstrained agents can be set to $+\infty$. Furthermore, we assume that $\myG$ is connected, as we can handle each connected component independently.

\subsection{Augmenting singleton constraints\texorpdfstring{ -- forcing $\mysize{V_i} \ge 2$}{}}\label{ssec:trans-vi-ge-2}

\emph{The transformation.}
For each constraint $i\in I$ with $|V_i|=1$, introduce three new agents, $s$, $t$, and $u$, two new objectives, $h$ and $\ell$, and one new constraint, $j$. Let $v \in V_i$ be the original agent adjacent to $i$, and let $k \in K_v$ be an objective adjacent to $v$. Set
$a_{is}=a_{jt}=a_{ju}=1$,
$c_{hs}=c_{\ell s}=1$, and
$c_{ht}=c_{\ell u}=2\sum_{w\in V_k}c_{kw}\min_{i\in I_w}a_{iw}^{-1}$.

\emph{Mapping back.}
Let $\myv{x}'$ be any feasible solution of the transformed instance. We map this back by setting $x_v=x_v'$ for all original agents $v\in V$.

\emph{Approximation ratio.}
Observe that we can always set $x_s'=0$ and $x_t'=x_u'=1/2$ without decreasing the objective value $\omega'(\myv{x}')$. Thus the optima of the original and transformed instances coincide, and any approximation ratio is preserved.

\subsection{Reducing the degree of constraints\texorpdfstring{ -- forcing $\mysize{V_i} = 2$}{}}\label{ssec:trans-vi-eq-2}

\emph{The transformation} \cite{floreen08local}.
Replace each constraint $i\in I$ with $|V_i|>2$ by the $\binom{|V_i|}{2}$ constraints
\begin{equation}\label{eq:transformed-constraints}
    a_{iu}x_u+a_{iv}x_v\leq 1,\quad u,v\in V_i,\ \ u<v.
\end{equation}

\emph{Mapping back.}
Let $\myv{x}'$ be an arbitrary feasible solution of the transformed instance. We map this back to a feasible solution $\myv{x}$ of the original instance by setting
\begin{equation}\label{eq:back-trans}
    x_v = \frac{2x_v'}{\max_{i\in I_v} |V_i|},\quad v\in V.
\end{equation}

To verify that $\myv{x}$ is feasible, consider an arbitrary original constraint $i\in I$. By the previous transformation, we have $|V_i|\geq 2$. Taking the sum over all the constraints \eqref{eq:transformed-constraints} replacing $i$, or, if $|V_i|=2$, considering the original constraint, we have
\[
    \textstyle
    \sum_{v\in V_i}\bigl(|V_i|-1\bigr)a_{iv}x_v'
    \le |V_i|\bigl(|V_i|-1\bigr) / 2.
\]
By \eqref{eq:back-trans} we thus have
\[
    \textstyle
    \sum_{v\in V_i}a_{iv}x_v
    \le \sum_{v\in V_i} 2a_{iv}x_v' / |V_i|
    \le 1.
\]

\emph{Approximation ratio.}
Because the objectives are unchanged in the transformation, we have $\omega(\myv{x})\geq 2\omega'(\myv{x}')/\Delta_I$ by linearity and \eqref{eq:back-trans}. Furthermore, an optimal solution of the original instance is a feasible solution of the transformed instance. Therefore, if $\myv{x}'$ is an $\alpha$-approximate solution, then $\myv{x}$ is a $\alpha\Delta_I/2$-approximate solution.

\subsection{Associating a unique objective with each agent\texorpdfstring{ -- forcing $\mysize{K_v} = 1$}{}}\label{ssec:trans-kv-eq-1}

\emph{The transformation.}
For each agent $v\in V$ with $|K_v|>1$, replace $v$ with $|K_v|$ copies of $v$ as follows. Associate each copy of $v$ with a unique objective in $K_v$. Replace each constraint adjacent to $v$ with $|K_v|$ copies of the constraint, with $v$ replaced by a unique copy in each constraint. The coefficients are unchanged.

\emph{Mapping back.}
Let $\myv{x}'$ be a feasible solution of the transformed instance. By symmetry we can assume that all copies $u$ of $v$ have the same value $x_{u}'$ without decreasing the objective value $\omega'(\myv{x}')$; indeed, if the values are different, just set all copies to the maximum value. Mapping back is done simply by identifying the copies back to the original.

\emph{Approximation ratio.}
Preserved. The optima of the original and the transformed instance coincide.

\subsection{Augmenting singleton objectives\texorpdfstring{ -- forcing $\mysize{V_k} \ge 2$}{}}\label{ssec:trans-vk-ge-2}

\emph{The transformation.}
For each objective $k\in K$ with $|V_k|=1$, let $v$ be the unique agent adjacent to $k$. Replace $v$ with two copies, $t$ and $u$, and replace each constraint adjacent to $v$ with two copies of the constraint, one containing $t$ in place of $v$, and the other containing $u$ in place of $v$. Let $c_{kt}=c_{ku}=c_{kv}/2$. The coefficients are otherwise unchanged.

\emph{Mapping back.}
Let $\myv{x}'$ be a feasible solution of the transformed instance. By symmetry we can assume that the copies of $v$ have the same value $x_{t}'=x_{u}'$ without decreasing the objective value $\omega'(\myv{x}')$; indeed, if the values are different, just set all copies to the maximum value. Mapping back is done simply by identifying the copies back to the original.

\emph{Approximation ratio.}
Preserved. The optima of the original and the transformed instance coincide.

\subsection{Normalising coefficients\texorpdfstring{ -- forcing $c_{kv} = 1$ for adjacent $k \in K$ and $v \in V$}{}}\label{ssec:trans-c01}

For each $v \in V$, let $k(v)$ be the unique objective in $K_v$.

\emph{The transformation.}
For each $v \in V$, $i \in I$, $k \in K$, divide $a_{iv}$ and $c_{kv}$ by $c_{k(v) v}$.

\emph{Mapping back.}
For each $v \in V$, multiply $x_v$ by $c_{k(v) v}$.

\emph{Approximation ratio.}
Preserved.

\section{Local approximation algorithm}\label{sec:approx}

Throughout this section we consider a max-min LP associated with a bipartite graph $\myG=(V\cup I\cup K,E)$ with these properties that follow from \S\ref{sec:unfolding} and \S\ref{sec:local-trans}:
\begin{enumerate}
    \item the graph $\myG$ is an unfolding of a finite graph,
    \item every agent $v\in V$ is adjacent to exactly one objective and at least one constraint,
    \item every constraint $i\in I$ is adjacent to exactly two agents,
    \item every objective $k\in K$ is adjacent to at least two agents,
    \item for adjacent $k \in K$ and $v \in V$, the coefficient $c_{kv}$ equals $1$.
\end{enumerate}
It follows that $\myG$ is countably infinite.

Recall that $k(v)$ is the unique objective adjacent to $v \in V$. Let $\myotherVK{v} = V_{k(v)} \setminus \{v\}$ be the set of other agents adjacent to this objective. For a constraint $i\in I$ and an agent $v\in V_i$, denote by $\myotherVI{v}{i}$ the unique agent other than $v$ in $V_i$. See Figure~\ref{fig:layers} for an illustration.

Let $R=2,3,\ldots$ be a fixed parameter that will determine the local horizon and the approximation ratio. Let $r=R-2$.

\subsection{An upper bound via alternating trees}\label{ssec:A_u}

A walk $W=(u_0,e_1,u_1,\ldots,e_\ell,u_\ell)$ in $\myG$ is \emph{alternating} if
(i) for all $1\leq j<j'\leq \ell$ with $u_j\in K$ and $u_{j'}\in K$, there exists a $j<j''<j'$ with $u_{j''}\in I$; and
(ii) for all $1\leq j<j'\leq \ell$ with $u_j\in I$ and $u_{j'}\in I$, there exists a $j<j''<j'$ with $u_{j''}\in K$.

Let $u\in V$ be an arbitrary agent, and consider the subgraph $\myA_u$ of $\myG$ induced by the nodes reachable  via alternating paths starting at $u$ that
(i) traverse the constraint $k(u)$ and have length at most $4r+3$; or
(ii) have length at most $1$.
The \emph{level} of a node of $\myA_u$ is its distance to $k(u)$, with the exception of $u$, which we define to have level $-1$, and the constraints adjacent to $u$, which we define to have level $-2$. See Figure~\ref{fig:layers} for an illustration.

Associate with $\myA_u$ a max-min LP by restriction from the max-min LP associated with~$\myG$.

\begin{lemma}\label{lem:au-structure}
    The graph $\myA_u$ is a finite tree. Moreover,
    \begin{enumerate}
    \item every objective in $\myA_u$ is at level\/ $0\pmod 4$,
    \item every agent in $\myA_u$ is at level either\/ $1\pmod 4$ or\/ $3\pmod 4$,
    \item every constraint in $\myA_u$ is at level\/ $2\pmod 4$,
    \item the leaves of $\myA_u$ are constraints at levels\/ $-2$ and\/ $4r+2$,
    \item for any objective $k$ in $\myA_u$ and any agent $v$ adjacent to $k$ in $\myG$, the agent $v$, occurs in $\myA_u$ and is adjacent to $k$ in $\myA_u$.
    \end{enumerate}
\end{lemma}
\begin{proof}
    By induction on $R$ using the assumptions on the structure of $\myG$.
\end{proof}

\begin{lemma}\label{lem:au-upper-bound}
    The optimum value of the max-min LP associated with $\myA_u$ is an upper bound on the value of any feasible solution of the max-min LP associated with $\myG$.
\end{lemma}
\begin{proof}
    Because $\myA_u$ is finite, the optimum is well defined. It suffices to show that any feasible solution of $\myG$ is (by restriction) a feasible solution of $\myA_u$. This follows from Lemma~\ref{lem:au-structure}: the objectives in $\myA_u$ are identical to those in $\myG$, and the constraints on variables in $\myA_u$ are either identical or relaxed from $\myG$ (at leaves or non-alternating constraints).
\end{proof}

\subsection{The optimum of \texorpdfstring{$\myA_u$}{Au}}\label{ssec:t_u}

The tree structure of $\myA_u$ enables a recursive 
characterisation of the optimum that proceeds
level-wise towards $u$.
Denote by $L(u,\ell)$ the set of all nodes at level
$\ell$ in $\myA_u$.

Let $t_u$ be the maximum value $\omega\geq 0$ such
that for all $d=0,1,\ldots,r$ it holds that the values
\begin{alignat}{2}
    \label{eq:au-recursion1}
    f_{u,v,0}^+(\omega) &= \min_{i\in I_v} a_{iv}^{-1}, &
    &\text{$v\in L\bigl(u,4r+1\bigr)$,} \\
    \label{eq:au-recursion2}
    f_{u,v,d}^-(\omega) &= \max\bigl(0,\, \omega - \textstyle\sum_{w \in \myotherVK{v}} f_{u,w,d}^+(\omega)\bigr), &
    &\text{$v\in L\bigl(u,4(r-d)-1\bigr)$,}\\
    \label{eq:au-recursion3}
    f_{u,v,d}^+(\omega) &= \min_{i\in I_v} a_{iv}^{-1}\bigl(1-\myaspacing_{i,\myotherVI{v}{i}}f_{u,\myotherVI{v}{i},d-1}^-(\omega)\bigr), &
    d\geq 1,\quad & \text{$v\in L\bigl(u,4(r-d)+1\bigr)$}\\
\intertext{satisfy the constraints}
    \label{eq:au-constraint1}
    f_{u,v,d}^+(\omega) &\geq 0, &
    0\leq d\leq r,\quad & \text{$v\in L\bigl(u,4(r-d)+1\bigr)$,}\\
    \label{eq:au-constraint2}
    f_{u,u,r}^-(\omega) &\leq \min_{i\in I_u}a_{iu}^{-1}. &
    &
\end{alignat}
Note that the maximum exists because \eqref{eq:au-recursion2} and \eqref{eq:au-recursion3} can be expressed using linear inequalities (by introducing additional variables), and $\omega=0$ is a feasible value. In a practical implementation of our algorithm, we do not need to invoke an LP solver; a simple binary search for an \emph{approximation} of $t_u$ is sufficient.

\begin{lemma}\label{lem:au-opt}
    Let $\myv{x}$ be any feasible solution that achieves the 
    objective value $\omega$ in the max-min LP associated with $\myA_u$.
    Then, for all $d=0,1,\ldots,r$ it holds that 
    \begin{alignat}{2}
    \label{eq:f-lb}
    0&\leq x_v\leq f_{u,v,d}^+(\omega)&\quad\text{for all }v&\in L\bigl(u,4(r-d)+1\bigr)\text{; and}\\
    \label{eq:f-ub}
    f_{u,v,d}^-(\omega)&\leq x_v\leq\min_{i\in I_v}a_{iv}^{-1}&\quad\text{for all }v&\in L\bigl(u,4(r-d)-1\bigr).
    \end{alignat}
    In particular, $t_u$ is the optimum utility of the max-min 
    LP associated with $\myA_u$.
\end{lemma}
\begin{proof}
    By induction on $d$ and in order of evaluation of the recursive steps \eqref{eq:au-recursion2} and \eqref{eq:au-recursion3}; see Appendix~\ref{app:lem:au-opt} for a full proof.
\end{proof}

In what follows we use the shorthand notation
$f_{u,v,d}^+=f_{u,v,d}^+(t_u)$
and
$f_{u,v,d}^-=f_{u,v,d}^-(t_u)$.

\subsection{Smoothing}\label{ssec:g}

For each agent $v\in V$ in $\myG$, let
$s_v$ be the minimum of the values $t_u$ over all
agents $u\in V$ at distance at most $4r+2$ from $v$ in $\myG$.
For all $v\in V$ in $\myG$ and all $d=0,1,\ldots,r$, define
\begin{alignat}{2}
    \label{eq:g-recursion1}
    g_{v,0}^+ &= \min_{i\in I_v} a_{iv}^{-1},&
    &\\
    \label{eq:g-recursion2}
    g_{v,d}^- &= \max\bigl(0,\, s_v-\textstyle\sum_{w \in \myotherVK{v}} g_{w,d}^+\bigr),&
    & \\
    \label{eq:g-recursion3}
    g_{v,d}^+ &= \min_{i\in I_v} a_{iv}^{-1}\bigl(1-\myaspacing_{i,\myotherVI{v}{i}}g_{\myotherVI{v}{i},d-1}^-\bigr),&
    \quad & d\geq1.
\end{alignat}

\begin{lemma}\label{lem:fg-ineq}
    For all $u\in V$ and all $d=0,1,\ldots,r$ it holds that
    \begin{alignat}{2}
        \label{eq:fg-ineq1}
        g_{v,d}^- &\leq f_{u,v,d}^- &
        \quad\text{ for all } v&\in L(u,4(r-d)-1) , \\
        \label{eq:fg-ineq2}
        f_{u,v,d}^+ &\leq g_{v,d}^+
        &\quad\text{ for all } v&\in L(u,4(r-d)+1) .
    \end{alignat}
\end{lemma}
\begin{proof}
    By induction on $d$, using the definition of $s_v$; see Appendix~\ref{app:lem:fg-ineq} for a full proof.
\end{proof}

\begin{lemma}\label{lem:g-bounds}
    For all $v\in V$ it holds that $g_{v,r}^+\geq 0$ and $g_{v,r}^-\leq\min_{i\in I_v}a_{iv}^{-1}$.
\end{lemma}
\begin{proof}
    Let $v \in V$ be arbitrary and choose $u \in \myotherVK{v}$; such a $u$ exists because every objective is adjacent to at least two agents. We have $g_{v,r}^+\geq f_{u,v,r}^+\geq 0$ by \eqref{eq:fg-ineq2} and \eqref{eq:au-constraint1}. Similarly, let $u=v$ to obtain
    \[
        g_{v,r}^-\leq f_{v,v,r}^-\leq \min_{i\in I_v}a_{iv}^{-1}
    \]
    by \eqref{eq:fg-ineq1} and \eqref{eq:au-constraint2}.
\end{proof}

\begin{lemma}\label{lem:g-monotone}
    For all $v\in V$ and $d=1,2,\ldots,r$ it holds that $g_{v,d-1}^-\leq g_{v,d}^-$ and $g_{v,d}^+\leq g_{v,d-1}^+$.
\end{lemma}
\begin{proof}
    By induction on $d$; see Appendix~\ref{app:lem:g-monotone} for a full proof.
\end{proof}

\begin{lemma}\label{lem:g-nonnegative}
    For all $v\in V$ and $d=0,1,\ldots,r$ it holds that $g_{v,d}^+\geq 0$.
\end{lemma}
\begin{proof}
    By Lemma~\ref{lem:g-bounds} and Lemma~\ref{lem:g-monotone}.
\end{proof}

Finally, each agent $v$ outputs the value
\begin{equation}\label{eq:x-def}
    x_v = \frac{1}{2R}\sum_{d=0}^r \bigl(g_{v,d}^++g_{v,d}^-\bigr).
\end{equation}
This completes the description of the algorithm. The algorithm is local, with the local horizon $\Theta(R)$. We proceed to show the vector $\myv{x}$ is a feasible solution, and within factor ${2 (1 - 1/\Dk)} + \epsilon'$ of the optimum.

\section{Analysis}\label{sec:analysis}

We start by partitioning the set of agents $V$ into \emph{up-agents} and \emph{down-agents} such that (i)~every constraint is adjacent to exactly one up-agent and exactly one down-agent; and (ii)~every objective is adjacent to exactly one up-agent.

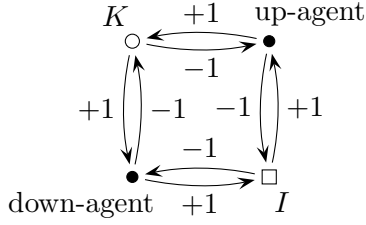
\begin{figure}
    \centering
    \input{fig-weights.tex}
    \caption{The weights used to assign the layers.}
    \label{fig:weights}
\end{figure}

Associate an integer \emph{layer} to each node of $\myG$ as follows. First, fix an arbitrary objective $k\in K$ to be at layer 0. Then, determine the layer of every other node $u$ by considering the unique directed path connecting $k$ to $u$. The layer of $u$ is determined by taking the sum of the weights of the directed edges in the path, where the weights are displayed in Figure~\ref{fig:weights}. Figure~\ref{fig:layers} displays an example of a partition into up- and down-agents together with an assignment of layers.

\begin{lemma}\label{lem:layer}
    The layers of the nodes of $\myG$ satisfy
    the following four properties:
    \begin{enumerate}
        \item every objective has layer\/ $0\pmod 4$,
        \item every down-agent has layer\/ $1\pmod 4$,
        \item every constraint has layer\/ $2\pmod 4$,
        \item every up-agent has layer\/ $3\pmod 4$.
    \end{enumerate}
\end{lemma}
\begin{proof}
    Immediate from Figure~\ref{fig:weights}.
\end{proof}

\subsection{Shifting strategy}\label{ssec:shifting}

Let $j=0,1,\ldots,R-1$ be a shift parameter.
For each agent $v\in V$, represent the layer of $v$
uniquely as $4(Rc+j)+4d+e$ for integers $c,d,e$
with $0\leq d\leq R-1$ and $e\in\{-1,1\}$.
Recall that $r=R-2$. Associate with $v$ the value 
\begin{equation}
\label{eq:yj-def}
y_v(j)=
\begin{cases}
  0 & \text{if $d=R-1$;} \\
  g_{v,r-d}^- & \text{if $d\leq R-2$ and $e=-1$;}\\
  g_{v,r-d}^+ & \text{if $d\leq R-2$ and $e=1$.}\\
\end{cases}
\end{equation}
Observe that a down-agent has $e = 1$ and an up-agent has $e = -1$, regardless of $j$.

For an objective $k$ and a vector $\myv{z}$ indexed by the agents, let $\omega_k(\myv{z}) = \sum_{v \in V_k} z_v$.

\begin{lemma}\label{lem:yj-analysis}
    The vector $\myv{y}(j)$ is a feasible solution of the max-min LP associated with $\myG$. For every objective $k \in K$, it holds that $\omega_k(\myv{y}(j)) = 0$ if $k$ is at layer\/ $4j-4\ (\mathrm{mod}\ 4R)$, and $\omega_k(\myv{y}(j)) \ge \min_{v\in V_k} s_v$ otherwise.
\end{lemma}
\begin{proof}
    From \eqref{eq:g-recursion1}, \eqref{eq:g-recursion2}, \eqref{eq:g-recursion3}, and \eqref{eq:yj-def}; see Appendix~\ref{app:lem:yj-analysis} for a full proof.
\end{proof}

Let us now average over all values of the shift parameter $j=0,1,\ldots,R-1$ to obtain
\begin{equation}\label{eq:y-def}
    y_v
    = \frac{1}{R}\sum_{j=0}^{R-1} y_v(j)
    = \begin{cases}
        \frac{1}{R}\sum_{d=0}^r g_{v,d}^- & \text{if $v$ is an up-agent}\\
        \frac{1}{R}\sum_{d=0}^r g_{v,d}^+ & \text{if $v$ is a down-agent.}
    \end{cases}
\end{equation}
\begin{lemma}\label{lem:y-analysis}
    The vector $\myv{y}$ is a feasible solution of the max-min LP associated with $\myG$. For every objective $k \in K$, it holds that $\omega_k(\myv{y}) \ge (1-1/R) \min_{v\in V_k} s_v$.
\end{lemma}
\begin{proof}
    Follows from Lemma~\ref{lem:yj-analysis}.
\end{proof}

\subsection{Averaging}\label{ssec:averaging}

Associate with each agent $v \in V$ a solution $\myv{y}^{\uparrow v}$ defined as follows. Choose the layers so that $v$ is an up-agent; this is always possible. Let $\myv{y}^{\uparrow v}$ be the value of \eqref{eq:y-def}.

\begin{lemma}\label{lem:x-feasible}
    The vector $\myv{x}$ is a feasible solution of the max-min LP associated with $\myG$.
\end{lemma}
\begin{proof}
    Consider an arbitrary constraint $i\in I$; let $V_i = \{v,w\}$. Note that whenever $v$ is an up-agent $w$ is a down-agent and vice versa. Let
    \[
        \myv{z} = \bigl(\myv{y}^{\uparrow v} + \myv{y}^{\uparrow w}\bigr) / 2.
    \]
    By \eqref{eq:y-def} and \eqref{eq:x-def} we have
    \[
        z_v
        = \frac{y^{\uparrow v}_v + y^{\uparrow w}_v}{2}
        = \frac{1}{2} \biggl(\frac{1}{R}\sum_{d=0}^r g_{v,d}^- + \frac{1}{R}\sum_{d=0}^r g_{v,d}^+ \biggr)
        = x_v
    \]
    and $z_w = x_w$. By Lemma~\ref{lem:y-analysis}, the solutions $\myv{y}^{\uparrow v}$ and $\myv{y}^{\uparrow w}$ do not violate the constraint $i$. Therefore
    \[
        a_v x_v + a_w x_w = a_v z_v + a_w z_w = \bigl((\myaspacing_v y^{\uparrow v}_v + \myaspacing_w y^{\uparrow v}_w) + (\myaspacing_v y^{\uparrow w}_v + \myaspacing_w y^{\uparrow w}_w)\bigr)/2 \le (1 + 1)/2 = 1.
    \]
    We conclude that the solution $\myv{x}$ does not violate the constraint $i$.
\end{proof}

\begin{lemma}\label{lem:x-objective}
    For every objective $k \in K$,
    \begin{equation}\label{eq:x-objective}
        \omega_k(\myv{x})
        \,\ge\, \frac{1}{2} \biggl(1-\frac{1}{R}\biggr) \frac{|V_k|}{|V_k|-1} \min_{v\in V_k} s_v.
    \end{equation}
\end{lemma}
\begin{proof}
    Consider an arbitrary objective $k$ in $\myG$. Note that whenever $v \in V_k$ is an up-agent, each $w \in \myotherVK{v}$ is a down-agent. Let
    \[
        \myv{z} = \frac{1}{|V_k|}\sum_{v\in V_k} \myv{y}^{\uparrow v}.
    \]
    By \eqref{eq:y-def} and \eqref{eq:x-def} we have
    \[
        \frac{|V_k|}{2(|V_k|-1)} z_u
        = \frac{1}{2(|V_k| - 1)} \sum_{v\in V_k} y^{\uparrow v}_u
        = \frac{1}{|V_k| - 1} \biggl(\frac{1}{2R}\sum_{d=0}^r g_{u,d}^- + \frac{\mysize{V_k} - 1}{2R}\sum_{d=0}^r g_{u,d}^+ \biggr)
        \le x_u
    \]
    for all $u \in V_k$. By Lemma~\ref{lem:y-analysis},
    \[
        \omega_k(\myv{x})
        = \sum_{u \in V_k} x_u
        \ge \sum_{u \in V_k} \frac{|V_k|}{2(|V_k|-1)} z_u
        \ge \frac{|V_k|}{2(|V_k|-1)} \biggl(1-\frac{1}{R}\biggr) \min_{v\in V_k} s_v .
    \]
    The claim follows.
\end{proof}

\subsection{Completing the analysis}

Lemmata \ref{lem:au-upper-bound} and \ref{lem:au-opt} show that for any $v \in V$, the value $t_v$ is an upper bound for the utility of any feasible solution of the max-min LP instance associated with $\myG$, and so is $s_v$. Lemma~\ref{lem:x-objective} therefore shows that our local approximation algorithm achieves the approximation ratio of ${2 (1 - 1/\Dk) (1 + 1/(R-1))}$, for the special case studied in \S\ref{sec:approx}.

Together with the local transformations of \S\ref{sec:local-trans}, taking into account the increase of the approximation in \S\ref{ssec:trans-vi-eq-2}, we conclude that the max-min LP problem admits a local algorithm with the approximation ratio of ${\Di (1 - 1/\Dk) (1 + 1/(R-1))}$; the local horizon is $\Theta(R)$. Theorem~\ref{thm:approx} follows by choosing a sufficiently large $R$. In particular, the constants $\Di$ and $\Dk$ are not required to run the algorithm.

\providecommand{\noopsort}[1]{}

\newpage\appendix
\section{}

\subsection{Local computability of the transformations in \S\ref{sec:local-trans}}\label{app:trans-comp}

\enlargethispage{1pt}

Even though a description of a local algorithm often involves interleaved steps of communication and computation, it should be noted that any local algorithm with local horizon $D$ can always be implemented as follows:
\begin{enumerate}
    \item Each node gathers full information about its radius $D$ neighbourhood; this is the \emph{local view} of the agent.
    \item Each node simulates the algorithm in its local view to determine its output.
\end{enumerate}
As pointed out in \S\ref{sec:unfolding}, we can assume that the local view is a tree. Furthermore, in our case it is sufficient that each \emph{agent} performs these steps -- constraints and objectives do not need to produce any output.

Therefore we can implement each transformation presented in this section as follows (with a small increase of the local horizon):
\begin{enumerate}
    \item Each agent gathers its local view, up to some constant distance. This is a tree.
    \item Each agent performs the transformation in its local view. The result is a graph, possibly with cycles.
    \item Each agent unfolds the graph to obtain a tree, discarding parts that are beyond its local horizon.
    \item Each agent simulates the rest of the local algorithm in this tree, and applies the back-mapping to determine its output.
\end{enumerate}
In some of the transformations, new agent nodes are created. In \S\ref{ssec:trans-vi-ge-2}, the output of the new agents is not needed. In \S\ref{ssec:trans-kv-eq-1} and \S\ref{ssec:trans-vk-ge-2}, existing agents can simulate their copies and compute the back-mapping.

The nontrivial part is to make sure that the transformations can be performed deterministically: if the local views of agents $u$ and $v$ partially overlap, and both agents perform a transformation in the common part, the common parts must be identical after the transformation. In particular, the port numbers must be identical. In the following, we show how to achieve this.

\S\ref{ssec:trans-vi-ge-2}:
Node $k$ is chosen to be the node in $K_v$ that has the smallest port number in $v$. The port number in $i$ for $\{i,s\}$ is one larger than the port number for $\{i,v\}$. Within the gadget defined by $s,t,u,h,\ell,j$, we can choose some fixed port numbering.

\S\ref{ssec:trans-vi-eq-2}, \S\ref{ssec:trans-kv-eq-1}, \S\ref{ssec:trans-vk-ge-2}:
These transformations can be summarised as follows: pick a subgraph, take $n$ copies of it, and discard some edges (e.g., in \S\ref{ssec:trans-kv-eq-1}, we take copies of a subgraph induced by an agent $v \in V$ and the adjacent constraints $I_v$). We can choose the port numbers of the transformed instance deterministically as follows. While we are taking copies of subgraphs, copy the port numbers as well. This may create duplicates. However, in each of these transformations, a set of duplicate port numbers in a node corresponds to a set of copies of one subgraph (e.g., in \S\ref{ssec:trans-kv-eq-1}, each copy of the subgraph induced by $v$ and $I_v$ creates exactly one duplicate port number for each agent $u \ne v$ adjacent to any $i \in I_v$). Furthermore, we can impose an ordering for the copies: in \S\ref{ssec:trans-vi-eq-2} by using pairs of port number in $i$; in \S\ref{ssec:trans-kv-eq-1} by using port numbers in $v$; and in \S\ref{ssec:trans-vk-ge-2} by arbitrarily ordering otherwise identical copies. Therefore we can re-number the ports in each node: first, order the ports by existing port numbers, using the ordering of the copies to break the ties; second, assign new port numbers $1, 2, \dotsc$ in this order.

\S\ref{ssec:trans-c01}:
The graph is not changed, and port numbering is preserved.

\subsection{Proof of Lemma~\ref{lem:au-opt}}\label{app:lem:au-opt}

To set up the base case at $d=0$, observe that by the feasibility of $\myv{x}$ and \eqref{eq:au-recursion1} we have
\[
    x_v
    \,\le\, \min_{i\in I_v} a_{iv}^{-1}
    \,=\, f_{u,v,0}^+(\omega)
    \qquad \text{for all } v\in L(u,4r+1) .
\]
Next consider any $0\leq d\leq r$ and assume inductively that 
\[
    x_w
    \,\le\, f_{u,w,d}^+(\omega)
    \qquad\text{for all } w\in L\bigl(u,4(r-d)+1\bigr) .
\]
Consider an arbitrary $v\in L(u,4(r-d)-1)$. Observe that for all $w\in V_{k(v)}$ it holds that either $v=w$ or $w\in L(u,4(r-d)+1)$. If we have
\[
    0
    \,\ge\, \omega - \!\!\!\sum_{w \in \myotherVK{v}}\!\! f_{u,w,d}^+(\omega),
\]
then \eqref{eq:au-recursion2} implies $f_{u,v,d}^-(\omega) = 0 \leq x_v$; otherwise \eqref{eq:au-recursion2} implies
\[
    f_{u,v,d}^-(\omega)
    \,=\, \omega - \!\!\!\sum_{w \in \myotherVK{v}}\!\! f_{u,w,d}^+(\omega)
    \,\le\, \omega - \!\!\!\sum_{w \in \myotherVK{v}}\!\! x_w
    \,\le\, x_v.
\]
Here the first inequality follows by the induction hypothesis, and the second inequality follows by assumption that $\myv{x}$ achieves the objective value $\omega$; in particular,
\[
    \sum_{w \in V_{k(v)}}\!\! x_w
    \,=\, x_v + \!\!\!\sum_{w \in \myotherVK{v}}\!\! x_w
    \,\ge\, \omega.
\]

To complete the induction, consider any $1\leq d\leq r$ and assume inductively that
\[
    f_{u,w,d-1}^-(\omega)
    \,\le\, x_w
    \qquad\text{for all } w\in L\bigl(u,4(r-(d-1))-1\bigr) .
\]
Consider an arbitrary $v\in L(u,4(r-d)+1)$. Observe that 
$\myotherVI{v}{i}\in L(u,4(r-(d-1))-1)$.
Because $\myv{x}$ is feasible, we have
$a_{iv}x_v + a_{i,\myotherVI{v}{i}}x_{\myotherVI{v}{i}} \leq 1$
for all $i \in I_v$. Thus, the inductive hypothesis and 
\eqref{eq:au-recursion3} imply
\[
  x_v
  \,\le\, \min_{i\in I_v} a_{iv}^{-1}\Bigl(1-a_{i,\myotherVI{v}{i}}x_{\myotherVI{v}{i}}\Bigr)
  \,\le\, \min_{i\in I_v} a_{iv}^{-1}\Bigl(1-\myaspacing_{i,\myotherVI{v}{i}}f_{u,\myotherVI{v}{i},d-1}^-(\omega)\Bigr)
  \,=\, f_{u,v,d}^+(\omega).
\]

To conclude that $t_u$ is the optimum utility of the max-min LP associated with $\myA_u$, consider an optimal solution $\myv{x}$ with the objective value $\omega^{*}$. Observe that \eqref{eq:f-lb} and \eqref{eq:f-ub} imply \eqref{eq:au-constraint1} and \eqref{eq:au-constraint2}. Therefore $t_u \ge \omega^{*}$. Furthermore, $t_u > \omega^{*}$ would contradict with the assumption that $\myv{x}$ is optimal. \qed

\pagebreak

\subsection{Proof of Lemma~\ref{lem:fg-ineq}}\label{app:lem:fg-ineq}

Consider an arbitrary $u \in V$. For all $w\in V$ at distance at most $4r+2$ from $u$ we have, by the definition of $s_w$,
\begin{equation}\label{eq:st-ineq}
    0\leq s_w\leq t_u.
\end{equation}

We proceed by induction on $d$. To establish the base case at $d=0$, observe by \eqref{eq:au-recursion1} and \eqref{eq:g-recursion1} that $f_{u,v,0}^+=g_{v,0}^+$ for all $v\in {L(u,4r+1)}$.

Next consider any $0\leq d\leq r$ and assume inductively that
\[
    f_{u,w,d}^+
    \,\le\, g_{w,d}^+
    \qquad\text{for all } w\in L(u,4(r-d)+1) .
\]
Consider an arbitrary $v\in L(u,4(r-d)-1)$. Observe that for all $w\in V_{k(v)}$ either $w=v$ or $w\in {L(u,4(r-d)+1)}$. Apply \eqref{eq:g-recursion2}, \eqref{eq:st-ineq}, the inductive hypothesis, and \eqref{eq:au-recursion2} to obtain
\[
    g_{v,d}^-
    \,=\, \max\biggl(\!0,\, s_v - \!\!\!\sum_{w \in \myotherVK{v}}\!\! g_{w,d}^+\biggr)
    \,\le\, \max\biggl(\!0,\, t_u - \!\!\!\sum_{w \in \myotherVK{v}}\!\! f_{u,w,d}^+\biggr)
    \,=\, f_{u,v,d}^-.
\]

Finally, consider any $1\leq d\leq r$ and assume inductively that
\[
    g_{w,d-1}^-
    \,\le\, f_{u,w,d-1}^-
    \qquad\text{for all } w\in L(u,4(r-(d-1))-1) .
\]
Consider an arbitrary $v \in L(u,4(r-d)+1)$. Observe that for all $i\in I_v$ it holds that $\myotherVI{v}{i} \in {L(u,4(r-(d-1))-1)}$. Apply \eqref{eq:au-recursion3}, the inductive hypothesis, and \eqref{eq:g-recursion3}, to obtain
\[
    f_{u,v,d}^+
    \,=\, \min_{i\in I_v} a_{iv}^{-1}\Bigl(1-\myaspacing_{i,\myotherVI{v}{i}}f_{u,\myotherVI{v}{i},d-1}^-\Bigr)
    \,\le\, \min_{i\in I_v} a_{iv}^{-1}\Bigl(1-\myaspacing_{i,\myotherVI{v}{i}}g_{\myotherVI{v}{i},d-1}^-\Bigr)
    \,=\, g_{v,d}^+.
\]
The induction is now complete.
\qed

\subsection{Proof of Lemma~\ref{lem:g-monotone}}\label{app:lem:g-monotone}

To set up the base case at $d=1$, observe first that $g_{v,0}^- \geq 0$ by \eqref{eq:g-recursion2}. By \eqref{eq:g-recursion3} and \eqref{eq:g-recursion1} thus
\[
    g_{v,1}^+
    \,=\, \min_{i\in I_v} a_{iv}^{-1}\Bigl(1-\myaspacing_{i,\myotherVI{v}{i}}g_{\myotherVI{v}{i},0}^-\Bigr)
    \,\le\, \min_{i\in I_v} a_{iv}^{-1}
    \,=\, g_{v,0}^+.
\]
Next consider any $1\leq d\leq r$ and assume inductively that $g_{v,d}^+\leq g_{v,d-1}^+$. Apply \eqref{eq:g-recursion2} and the inductive hypothesis to obtain
\[
    g_{v,d}^-
    \,=\, \max\biggl(\!0,\, s_v - \!\!\!\sum_{w \in \myotherVK{v}}\!\! g_{w,d}^+\biggr)
    \,\ge\, \max\biggl(\!0,\, s_v - \!\!\!\sum_{w \in \myotherVK{v}}\!\! g_{w,d-1}^+\biggr)
    \,=\, g_{v,d-1}^- .
\]
Finally, consider any $2\leq d\leq r$ and assume inductively that $g_{v,d-2}^-\leq g_{v,d-1}^-$. Apply \eqref{eq:g-recursion3} and the inductive hypothesis to obtain
\[
    g_{v,d}^+
    \,=\, \min_{i\in I_v} a_{iv}^{-1}\Bigl(1-\myaspacing_{i,\myotherVI{v}{i}}g_{\myotherVI{v}{i},d-1}^-\Bigr)
    \,\le\, \min_{i\in I_v} a_{iv}^{-1}\Bigl(1-\myaspacing_{i,\myotherVI{v}{i}}g_{\myotherVI{v}{i},d-2}^-\Bigr)
    \,=\, g_{v,d-1}^+ .
\]
This completes the induction.
\qed

\subsection{Proof of Lemma~\ref{lem:yj-analysis}}\label{app:lem:yj-analysis}

\emph{Feasibility.}
The vector is nonnegative by \eqref{eq:g-recursion2} and Lemma~\ref{lem:g-nonnegative}. Consider an arbitrary constraint $i\in I$. By Lemma~\ref{lem:layer} we can represent the layer of $i$ uniquely as $4(Rc+j)+4d+2$ for integers $c,d$ with $0\leq d\leq R-1$. Let $V_i=\{v,w\}$ with $v$ at layer $4(Rc+j)+4d+1$ and $w$ at layer $4(Rc+j)+4(d+1)-1$.

First consider the case $d=R-1$. Note that the layer of $w$ is actually $4(R(c+1)+j)-1$, that is, $d=0$ and $e=-1$ for $w$. By \eqref{eq:yj-def} and Lemma~\ref{lem:g-bounds}, we have
\[
    a_{iv}y_v(j)+a_{iw}y_w(j)
    \,=\, a_{iw}g_{w,r}^-
    \,\le\, 1.
\]
Next consider the case $d=R-2$. By \eqref{eq:yj-def} and \eqref{eq:g-recursion1}, we have
\[
    a_{iv}y_v(j)+a_{iw}y_w(j)
    \,=\, a_{iv}g_{v,0}^+
    \,\le\, 1.
\]
Finally consider the case $d<R-2$. By \eqref{eq:yj-def} and \eqref{eq:g-recursion3}, we have
\[
\begin{split}
    a_{iv}y_v(j)+a_{iw}y_w(j)
    &\,=\, a_{iv}g_{v,r-d}^++a_{iw}g_{w,r-d-1}^- \\
    &\,\le\, a_{iv}a_{iv}^{-1}(1-a_{iw}g_{w,r-d-1}^-)+a_{iw}g_{w,r-d-1}^-
    \,=\, 1.
\end{split}
\]
The claim follows since $i$ was arbitrary.

\emph{Objectives.}
Consider an arbitrary objective $k$ in $\myG$. By Lemma~\ref{lem:layer} we can represent the layer of $k$ uniquely as $4(Rc+j)+4d$ for integers $c,d$ with $0\leq d\leq R-1$. There is a unique up-agent in $V_k$ at layer $4(Rc+j)+4d-1$. Denote this agent by $v$. The other agents $w \in \myotherVK{v}$ are down-agents at layer $4(Rc+j)+4d+1$.

First consider the case $d=R-1$. By \eqref{eq:yj-def} we have
\[
    \omega_k(\myv{y}(j))
    \,=\, y_v(j)+\sum_{w \in \myotherVK{v}} y_w(j)
    \,=\, 0.
\]
Then consider the case $d\leq R-2$. By \eqref{eq:yj-def} and \eqref{eq:g-recursion2}, we have
\[
\begin{split}
    \omega_k(\myv{y}(j))
    &\,=\, y_v(j) \,+\!\!\sum_{w \in \myotherVK{v}}\!\! y_w(j)
    \,=\, g_{v,r-d}^- \,+\, \!\!\!\sum_{w \in \myotherVK{v}}\!\! g_{w,r-d}^+\\
    &\,\ge\, s_v
                \,-\!\!\sum_{w \in \myotherVK{v}}\!\! g_{w,r-d}^+
                \,+\!\!\sum_{w \in \myotherVK{v}}\!\! g_{w,r-d}^+
    \,=\, s_v
    \,\ge\, \min_{u\in V_k} s_u.
\end{split}
\]
The claim follows because $k$ was arbitrary.
\qed

\end{document}

%% file: fig-settings.tex
\SpecialCoor

\psset{unit=1.0cm}

\definecolor{grid}{rgb}{0.8,0.8,1.0}
\definecolor{subgrid}{rgb}{0.9,0.9,1.0}
\definecolor{gridlabel}{rgb}{0.8,0.8,1.0}
\definecolor{mygrey}{rgb}{0.6,0.6,0.6}

\psset{%
    gridcolor=grid,
    subgridcolor=subgrid,
    gridlabelcolor=gridlabel,
    gridwidth=0.2pt,
    subgridwidth=0.1pt}

\newcommand{\mypsgrid}{%
}

\newcommand{\mypsnormal}{\psset{%
    fillstyle=none,
    fillcolor=black,
    linestyle=solid,
    linecolor=black,
    linewidth=0.4pt}}

\mypsnormal

\newcommand{\mypsK}{\mypsnormal\psset{%
    radius=1mm,
    labelsep=1.3mm,
    fillstyle=solid,
    fillcolor=white}}

\newcommand{\mypsKpassive}{\mypsK\psset{%
    linecolor=mygrey}}

\newcommand{\mypsI}{\mypsnormal\psset{%
    framesize=2mm,
    labelsep=1.5mm,
    fillstyle=solid,
    fillcolor=white}}

\newcommand{\mypsV}{\mypsnormal\psset{%
    radius=0.8mm,
    labelsep=1.5mm,
    fillstyle=solid,
    fillcolor=black}}

\newcommand{\mypsVpassive}{\mypsV\psset{%
    linecolor=mygrey,
    fillcolor=mygrey}}

\newcommand{\mypsE}{\mypsnormal}

\newcommand{\mypsEpassive}{\mypsE\psset{%
    linecolor=mygrey}}

\newcommand{\mypsEdashed}{\mypsE\psset{%
    linestyle=dashed,
    dash=1.5pt 1.5pt}}

\newcommand{\mypsEhighlight}{\mypsE\psset{%
    linewidth=2.0pt}}

%% file: fig-layers.tex
\newcommand{\mync}[2]{%
    \mypsE%
    \ncarc[arcangle=-5]{#1}{#2}%
}
\newcommand{\myncp}[2]{%
    \mypsEpassive%
    \ncarc[arcangle=-5]{#1}{#2}%
}
\newcommand{\mynch}[2]{%
    \mypsEhighlight%
    \ncarc[arcangle=-5]{#1}{#2}%
}
\newcommand{\mytextadjust}[1]{\raisebox{-3pt}{#1}}
\newcommand{\mytext}[2]{%
    \rput[Br]{0}(\curcol,#1){\mytextadjust{#2}}%
}
\newcommand{\myhead}[3]{%
    \rput[B#2]{0}(#1,4){#3}
}
\newcommand{\curcol}{}

\psset{xunit=10mm,yunit=\baselineskip}
\begin{pspicture}(-5.0,-11.4)(8.5,4.7)
    \mypsgrid%
    \qline(-5.0,3.6)(2.5,3.6)%
    \renewcommand{\curcol}{-1.5}%
    \myhead{\curcol}{r}{node type}%
    \mytext{3}{passive down-agents}%
    \mytext{2}{constraints}%
    \mytext{1}{up-agents}%
    \mytext{0}{objectives}%
    \mytext{-1}{down-agents}%
    \mytext{-2}{constraints}%
    \mytext{-3}{up-agents}%
    \mytext{-4}{objectives}%
    \mytext{-5}{down-agents}%
    \mytext{-6}{constraints}%
    \mytext{-7}{passive up-agents}%
    \mytext{-8}{passive objectives}%
    \mytext{-9}{passive down-agents}%
    \mytext{-10}{constraints}%
    \mytext{-11}{up-agents}%
    \renewcommand{\curcol}{-0.1}%
    \myhead{0.1}{r}{layer}%
    \multido{\ic=3+-1,\iv=-3+1}{15}{%
        \rput[Br]{0}(\curcol,\ic){\mytextadjust{$\iv$}}
    }%
    \renewcommand{\curcol}{1.5}%
    \myhead{2.5}{r}{level in $\myA_u$}%
    \multido{\ic=2+-1,\iv=-2+1}{9}{%
       \rput[Br]{0}(\curcol,\ic){\mytextadjust{$\iv$}}%
    }%
    \mypsVpassive%
    \Cnode(3.25,3){daa}%
    \Cnode(4,3){dab}%
    \Cnode(5,3){dac}%
    \Cnode(5.5,3){dad}%
    \Cnode(6.25,3){dae}%
    \nput{0}{dae}{$\myotherVI{v}{i}$}%
    \mypsI%
    \fnode(3.25,2){iba}%
    \fnode(4,2){ibb}%
    \fnode(5,2){ibc}%
    \fnode(5.5,2){ibd}%
    \fnode(6.25,2){ibe}%
    \nput{0}{ibe}{$i$}%
    \mypsV%
    \Cnode(3.25,1){uba}%
    \Cnode(4,1){ubb}%
    \Cnode(5.25,1){ubc}%
    \nput{180}{ubc}{$u$}%
    \Cnode(6.25,1){ubd}%
    \nput{0}{ubd}{$v$}%
    \mypsK%
    \Cnode(3.25,0){kba}%
    \Cnode(4,0){kbb}%
    \Cnode(5.25,0){kbc}%
    \Cnode(6.25,0){kbd}%
    \nput{0}{kbd}{$k(v)$}%
    \mypsV%
    \Cnode(3.25,-1){dba}%
    \Cnode(4,-1){dbb}%
    \Cnode(5,-1){dbc}%
    \Cnode(5.75,-1){dbd}%
    \Cnode(6.25,-1){dbe}%
    \nput{0}{dbe}{$w \in \myotherVK{v}$}%
    \mypsI%
    \fnode(3,-2){ica}%
    \fnode(3.5,-2){icb}%
    \fnode(4,-2){icc}%
    \fnode(5,-2){icd}%
    \fnode(5.75,-2){ice}%
    \fnode(6.25,-2){icf}%
    \mypsV%
    \Cnode(3,-3){uca}%
    \Cnode(3.5,-3){ucb}%
    \Cnode(4.5,-3){ucc}%
    \Cnode(6,-3){ucd}%
    \mypsK%
    \Cnode(3,-4){kca}%
    \Cnode(3.5,-4){kcb}%
    \Cnode(4.5,-4){kcc}%
    \Cnode(6,-4){kcd}%
    \mypsV%
    \Cnode(3,-5){dca}%
    \Cnode(3.5,-5){dcb}%
    \Cnode(4,-5){dcc}%
    \Cnode(5,-5){dcd}%
    \Cnode(6,-5){dce}%
    \mypsI%
    \fnode(3,-6){ida}%
    \fnode(3.5,-6){idb}%
    \fnode(4,-6){idc}%
    \fnode(4.5,-6){idd}%
    \fnode(5,-6){ide}%
    \fnode(5.5,-6){idf}%
    \fnode(6,-6){idg}%
    \mypsVpassive%
    \Cnode(3,-7){uda}%
    \Cnode(3.75,-7){udb}%
    \Cnode(4.5,-7){udc}%
    \Cnode(5,-7){udd}%
    \Cnode(5.5,-7){ude}%
    \Cnode(6,-7){udf}%
    \mypsKpassive%
    \Cnode(3,-8){kda}%
    \Cnode(3.75,-8){kdb}%
    \Cnode(4.5,-8){kdc}%
    \Cnode(5,-8){kdd}%
    \Cnode(5.5,-8){kde}%
    \Cnode(6,-8){kdf}%
    \mypsVpassive%
    \Cnode(3,-9){dda}%
    \Cnode(3.5,-9){ddb}%
    \Cnode(4,-9){ddc}%
    \Cnode(4.5,-9){ddd}%
    \Cnode(5,-9){dde}%
    \Cnode(5.5,-9){ddf}%
    \Cnode(6,-9){ddg}%
    \mypsI%
    \fnode(3,-10){iea}%
    \fnode(3.5,-10){ieb}%
    \fnode(4,-10){iec}%
    \fnode(4.5,-10){ied}%
    \fnode(5,-10){iee}%
    \fnode(5.5,-10){ief}%
    \fnode(6,-10){ieg}%
    \mypsV%
    \Cnode(3,-11){uea}%
    \Cnode(3.5,-11){ueb}%
    \Cnode(4,-11){uec}%
    \Cnode(4.5,-11){ued}%
    \Cnode(5,-11){uee}%
    \Cnode(5.5,-11){uef}%
    \Cnode(6,-11){ueg}%
    \myncp{daa}{iba}%
    \myncp{dab}{ibb}%
    \myncp{dac}{ibc}%
    \myncp{dad}{ibd}%
    \myncp{dae}{ibe}%
    \mync{iba}{uba}%
    \mync{ibb}{ubb}%
    \mynch{ibc}{ubc}%
    \mynch{ibd}{ubc}%
    \mync{ibe}{ubd}%
    \mync{uba}{kba}%
    \mync{ubb}{kbb}%
    \mynch{ubc}{kbc}%
    \mync{ubd}{kbd}%
    \mync{kba}{dba}%
    \mync{kbb}{dbb}%
    \mynch{kbc}{dbc}%
    \mynch{kbc}{dbd}%
    \mync{kbd}{dbe}%
    \mync{dba}{ica}%
    \mync{dba}{icb}%
    \mync{dbb}{icc}%
    \mynch{dbc}{icd}%
    \mynch{dbd}{ice}%
    \mync{dbe}{icf}%
    \mync{ica}{uca}%
    \mync{icb}{ucb}%
    \mync{icc}{ucc}%
    \mynch{icd}{ucc}%
    \mynch{ice}{ucd}%
    \mync{icf}{ucd}%
    \mync{uca}{kca}%
    \mync{ucb}{kcb}%
    \mynch{ucc}{kcc}%
    \mynch{ucd}{kcd}%
    \mync{kca}{dca}%
    \mync{kcb}{dcb}%
    \mynch{kcc}{dcc}%
    \mynch{kcc}{dcd}%
    \mynch{kcd}{dce}%
    \mync{dca}{ida}%
    \mync{dcb}{idb}%
    \mynch{dcc}{idc}%
    \mynch{dcd}{idd}%
    \mynch{dcd}{ide}%
    \mynch{dcd}{idf}%
    \mynch{dce}{idg}%
    \myncp{ida}{uda}%
    \myncp{idb}{udb}%
    \myncp{idc}{udb}%
    \myncp{idd}{udc}%
    \myncp{ide}{udd}%
    \myncp{idf}{ude}%
    \myncp{idg}{udf}%
    \myncp{uda}{kda}%
    \myncp{udb}{kdb}%
    \myncp{udc}{kdc}%
    \myncp{udd}{kdd}%
    \myncp{ude}{kde}%
    \myncp{udf}{kdf}%
    \myncp{kda}{dda}%
    \myncp{kdb}{ddb}%
    \myncp{kdb}{ddc}%
    \myncp{kdc}{ddd}%
    \myncp{kdd}{dde}%
    \myncp{kde}{ddf}%
    \myncp{kdf}{ddg}%
    \myncp{dda}{iea}%
    \myncp{ddb}{ieb}%
    \myncp{ddc}{iec}%
    \myncp{ddd}{ied}%
    \myncp{dde}{iee}%
    \myncp{ddf}{ief}%
    \myncp{ddg}{ieg}%
    \mync{iea}{uea}%
    \mync{ieb}{ueb}%
    \mync{iec}{uec}%
    \mync{ied}{ued}%
    \mync{iee}{uee}%
    \mync{ief}{uef}%
    \mync{ieg}{ueg}%
\end{pspicture}

%% file: fig-trans.tex
\newcommand{\mync}[2]{%
    \mypsE%
    \ncarc[arcangle=5]{#1}{#2}%
}
\newcommand{\myncx}[2]{%
    \mypsEdashed%
    \pcarc[arcangle=0.1]([angle=#2,nodesep=0pt]#1)([angle=#2,nodesep=0.5]#1)%
}
\newcommand{\mymapsto}[2]{%
    \rput{-90}(#1,3.25){$\longmapsto$}%
    \rput[r]{0}(#1,3.25){\S\ref{#2}\hspace{1ex}}%
}
\psset{unit=7mm}
\begin{pspicture}(-1.0,-0.6)(18,6.6)
    \mypsgrid%
    \mymapsto{1}{ssec:trans-vi-ge-2}%
    \mypsV\Cnode(0,5){AAv}\nput{90}{AAv}{$v$}%
    \mypsI\fnode(1,5){AAi}\nput{90}{AAi}{$i$}%
    \mync{AAv}{AAi}%
    \myncx{AAv}{180}%
    \mypsV\Cnode(0,1){ABv}\nput{90}{ABv}{$v$}%
    \mypsV\Cnode(2,1){ABs}\nput[labelsep=0.5mm]{135}{ABs}{$s$}%
    \mypsV\Cnode(3,2){ABt}\nput[labelsep=0.7mm]{45}{ABt}{$t$}%
    \mypsV\Cnode(3,0){ABu}\nput[labelsep=0.7mm]{-45}{ABu}{$u$}%
    \mypsI\fnode(1,1){ABi}\nput{90}{ABi}{$i$}%
    \mypsI\fnode(3,1){ABj}\nput{0}{ABj}{$j$}%
    \mypsK\Cnode(2,2){ABh}\nput[labelsep=0.7mm]{135}{ABh}{$h$}%
    \mypsK\Cnode(2,0){ABl}\nput[labelsep=1mm]{-150}{ABl}{$\ell$}%
    \mync{ABv}{ABi}%
    \mync{ABi}{ABs}%
    \mync{ABs}{ABh}%
    \mync{ABh}{ABt}%
    \mync{ABt}{ABj}%
    \mync{ABj}{ABu}%
    \mync{ABu}{ABl}%
    \mync{ABl}{ABs}%
    \myncx{ABv}{180}%
    \mymapsto{6}{ssec:trans-vi-eq-2}%
    \pnode(6,5){BAip}
    \mypsV\Cnode([angle=60,nodesep=1]BAip){BAv1}%
    \mypsV\Cnode([angle=180,nodesep=1]BAip){BAv2}%
    \mypsV\Cnode([angle=-60,nodesep=1]BAip){BAv3}%
    \mypsI\fnode(BAip){BAi}\nput{120}{BAi}{$i$}%
    \mync{BAi}{BAv1}%
    \mync{BAi}{BAv2}%
    \mync{BAi}{BAv3}%
    \myncx{BAv1}{60}%
    \myncx{BAv2}{180}%
    \myncx{BAv3}{-60}%
    \pnode(6,1){BBip}%
    \mypsV\Cnode([angle=60,nodesep=1]BBip){BBv1}%
    \mypsV\Cnode([angle=180,nodesep=1]BBip){BBv2}%
    \mypsV\Cnode([angle=-60,nodesep=1]BBip){BBv3}%
    \mypsI\fnode([angle=120,nodesep=0.5]BBip){BBi1}%
    \mypsI\fnode([angle=240,nodesep=0.5]BBip){BBi2}%
    \mypsI\fnode([angle=0,nodesep=0.5]BBip){BBi3}%
    \mync{BBi1}{BBv1}%
    \mync{BBv1}{BBi3}%
    \mync{BBi3}{BBv3}%
    \mync{BBv3}{BBi2}%
    \mync{BBi2}{BBv2}%
    \mync{BBv2}{BBi1}%
    \myncx{BBv1}{60}%
    \myncx{BBv2}{180}%
    \myncx{BBv3}{-60}%
    \mymapsto{11}{ssec:trans-kv-eq-1}%
    \mypsK\Cnode(9,4.25){CAk1}%
    \mypsK\Cnode(9,5.75){CAk2}%
    \mypsV\Cnode(10,5){CAv}\nput{90}{CAv}{$v$}%
    \mypsI\fnode(11,4.5){CAi1}%
    \mypsI\fnode(11,5.5){CAi2}%
    \mypsV\Cnode(12,4){CAv1}%
    \mypsV\Cnode(12,6){CAv2}%
    \mync{CAk1}{CAv}%
    \mync{CAk2}{CAv}%
    \mync{CAv}{CAi1}%
    \mync{CAv}{CAi2}%
    \mync{CAi1}{CAv1}%
    \mync{CAi2}{CAv2}%
    \myncx{CAk1}{180}%
    \myncx{CAk2}{180}%
    \myncx{CAv1}{0}%
    \myncx{CAv2}{0}%
    \mypsK\Cnode(9,0.25){CBk1}%
    \mypsK\Cnode(9,1.75){CBk2}%
    \mypsV\Cnode(10,0.25){CBva}%
    \mypsV\Cnode(10,1.75){CBvb}%
    \mypsI\fnode(11,0.0){CBi1a}%
    \mypsI\fnode(11,0.5){CBi2a}%
    \mypsI\fnode(11,1.5){CBi1b}%
    \mypsI\fnode(11,2.0){CBi2b}%
    \mypsV\Cnode(12,0){CBv1}%
    \mypsV\Cnode(12,2){CBv2}%
    \mync{CBk1}{CBva}%
    \mync{CBk2}{CBvb}%
    \mync{CBva}{CBi1a}%
    \mync{CBva}{CBi2a}%
    \mync{CBvb}{CBi1b}%
    \mync{CBvb}{CBi2b}%
    \mync{CBi1a}{CBv1}%
    \mync{CBi2a}{CBv2}%
    \mync{CBi1b}{CBv1}%
    \mync{CBi2b}{CBv2}%
    \myncx{CBk1}{180}%
    \myncx{CBk2}{180}%
    \myncx{CBv1}{0}%
    \myncx{CBv2}{0}%
    \mymapsto{16}{ssec:trans-vk-ge-2}%
    \mypsK\Cnode(14,5){DAk}\nput{90}{DAk}{$k$}%
    \mypsV\Cnode(15,5){DAv}\nput{90}{DAv}{$v$}%
    \mypsI\fnode(16,4.5){DAi1}%
    \mypsI\fnode(16,5.5){DAi2}%
    \mypsV\Cnode(17,4){DAv1}%
    \mypsV\Cnode(17,6){DAv2}%
    \mync{DAk}{DAv}%
    \mync{DAv}{DAi1}%
    \mync{DAv}{DAi2}%
    \mync{DAi1}{DAv1}%
    \mync{DAi2}{DAv2}%
    \myncx{DAv1}{0}%
    \myncx{DAv2}{0}%
    \mypsK\Cnode(14,1){DBk}\nput{90}{DBk}{$k$}%
    \mypsV\Cnode(15,0.25){DBu}\nput{90}{DBu}{$u$}%
    \mypsV\Cnode(15,1.75){DBt}\nput{90}{DBt}{$t$}%
    \mypsI\fnode(16,0.0){DBi1a}%
    \mypsI\fnode(16,0.5){DBi2a}%
    \mypsI\fnode(16,1.5){DBi1b}%
    \mypsI\fnode(16,2.0){DBi2b}%
    \mypsV\Cnode(17,0){DBv1}%
    \mypsV\Cnode(17,2){DBv2}%
    \mync{DBk}{DBu}%
    \mync{DBk}{DBt}%
    \mync{DBu}{DBi1a}%
    \mync{DBu}{DBi2a}%
    \mync{DBt}{DBi1b}%
    \mync{DBt}{DBi2b}%
    \mync{DBi1a}{DBv1}%
    \mync{DBi2a}{DBv2}%
    \mync{DBi1b}{DBv1}%
    \mync{DBi2b}{DBv2}%
    \myncx{DBv1}{0}%
    \myncx{DBv2}{0}%
\end{pspicture}

%% file: fig-weights.tex
\psset{unit=9mm}
\begin{pspicture}(-3.0,-1.6)(3.0,1.6)
\mypsgrid%
\mypsK%
\Cnode(-1,1){k}\nput{120}{k}{$K$}%
\mypsI%
\fnode(1,-1){i}\nput[labelsep=1mm]{-60}{i}{$I$}%
\mypsV%
\Cnode(1,1){up}\nput{60}{up}{up-agent}%
\Cnode(-1,-1){down}\nput{-120}{down}{down-agent}%
\mypsE%
\psset{arcangle=-12,arrows=->,arrowscale=2,nodesep=0.1,labelsep=1mm}%
\ncarc{down}{k}\nbput{$-1$}%
\ncarc{k}{up}\nbput{$-1$}%
\ncarc{up}{i}\nbput{$-1$}%
\ncarc{i}{down}\nbput{$-1$}%
\ncarc{k}{down}\nbput{$+1$}%
\ncarc{down}{i}\nbput{$+1$}%
\ncarc{i}{up}\nbput{$+1$}%
\ncarc{up}{k}\nbput{$+1$}%
\end{pspicture}